\documentstyle[preprint,eqsecnum,prd,aps,epsf]{revtex}
\tightenlines
\begin{document}
\draft
%\twocolumn[\hsize\textwidth\columnwidth\hsize\csname@twocolumnfalse\endcsname

\title{Photon Production of Axionic Cold Dark Matter}
%\title{Photon Production from Coherent Oscillations of Axionic Condensate}

\author{Da-Shin Lee\footnote{\tt E-mail address: dslee@cc3.ndhu.edu.tw}}
\address{Department of Physics, National Dong Hwa University, Hua-Lien, 
Taiwan, R.O.C.}

\author{Kin-Wang Ng\footnote{\tt E-mail address: nkw@phys.sinica.edu.tw}}
\address{Institute of Physics, Academia Sinica, Taipei, Taiwan, R.O.C.}

\date{September 1999}
\maketitle

\begin{abstract}
Using the non-equilibrium quantum field theory,  photon production from the 
coherently oscillating axion
field in a flat Robertson-Walker cosmology is re-examined.  
First neglecting the Debye screening of the baryon plasma to photons , 
we find that the axions will 
dissipate into photons via spinodal instability in addition to
parametric resonance.  As a result of the pseudo-scalar nature of the 
axion-photon coupling ,  we observe a circular polarization 
asymmetry in the produced  photons.   
However, these effects are suppressed to an insignificant level in the  
expanding universe. 
We then briefly discuss a systematic way of including the plasma effect 
which can further suppress the photon production. 
We note that the formalism of the problem can be applied to any pseudo-scalar 
field coupled to photon in a thermal background in a general curved 
spacetime.
\end{abstract}

\vspace{2pc}

\section{INTRODUCTION}
\label{INT}

It is compelling that most of the matter in the universe is in a form
of non-baryonic cold dark matter. If it exists, it would play an important
role in the structure formation of the universe~\cite{tur}. 
Axions, the  pseudo-Goldstone bosons, are among the most promising candidates for 
the non-baryonic cold dark matter. They arise from the spontaneous breaking of a 
global
$U(1)$ symmetry of Peccei and Quinn (PQ), which is introduced
to solve the strong CP problem of QCD~\cite{pre,kol,pec}. 
In the standard big-bang cosmology, 
after the spontaneous breakdown of the PQ symmetry,  the expectation value of the 
axion field (i.e. the axionic condensate)  takes some random value on the interval 
$[0,2\pi]$ and,  is  approximately constant over length scales which are 
smaller than the horizon size~\cite{vil}. 
If inflation occurs either after or during
the PQ symmetry breaking, then the expectation value can be nearly
constant throughout the entire universe~\cite{pi}. 
At high temperatures above the 
$\Lambda_{\rm QCD}$ scale, the axion is massless; however, at low temperatures,
the axion develops a mass due to QCD instanton effects~\cite{gro}. 
Once the axion mass becomes greater than the universe expansion rate, 
 the expectation value of the axion  field begins to oscillate coherently around 
the minimum of its effective potential that is near the origin. 
The  oscillating axion field then  dissipates mainly due to the universe expansion 
as well as particle production~\cite{pre,kol}.

In the original papers~\cite{pre}, simple estimates of  the thermal dissipation
of the homogeneous axionic condensate   were given. They considered instabilities 
arising from the parametric amplification of quantum fluctuations that  could 
 pump the energy of the homogeneous axionic condensate into  its quantum 
fluctuations via self couplings,
as well as into quantum fluctuating photon modes via a coupling of the axion to 
electromagnetism due to
the color anomaly of the PQ symmetry.
This  dissipational dynamics via quantum particle production  exhibits the feature 
of unstable bands, and an exponential growth of the quantum fluctuating modes that 
are characteristics of parametric resonance. 
The growth of the modes in the unstable bands translates 
into profuse particle production. 
A given unstable mode will grow as long as it lies within the unstable band.  
However, eventually it will be red-shifted out of 
the band as the universe expands, and then the instabilities of parametric 
resonance are shut off.    
In Ref.~\cite{pre}, it has been shown   that  for the PQ symmetry breaking 
scale $f_a > 10^{12}{\rm GeV}$,
 because the axion is very weakly 
coupled,  the time it takes  to be red-shifted out of the unstable band is  too 
short to build up an appreciable growth  of the quantum fluctuating modes. 
Thus, all of these effects are insignificant. The condensate is effectively 
nondissipative and pressureless.
It would survive in the expanding universe, and it behaves like cold dust
at the present time. Interestingly, if $f_a \sim 10^{12}{\rm GeV}$, it 
could constitute a major component of the dark matter of the universe.

Recently, the authors of Ref.~\cite{kol2} were
motivated by the recent understanding of the important role of the
spinodal instability and   parametric resonance  that provide the nonlinear and 
nonperturbative mechanisms in the quantum particle production 
driven by  the large amplitude oscillations of the coherent  field 
~\cite{boy1,boy2,boy3,boy4,lee}.  They
re-examined the issue of the dissipation of 
the  axion field resulting from the production of its  quantum fluctuations. They 
confirmed  that 
the presence of the parametric resonance would lead to an explosive 
growth of quantum fluctuations if the universe was Minkowskian. 
Taking  account of  the expansion of the  universe, quantum fluctuations of the 
axion do not 
become significant. This result confirms the conventional wisdom. 

In this paper, we will re-examine the  damping dynamics of the axion
  arising from photon production in an expanding universe 
in the context of the  non-equilibrium quantum field theory.
The goal of this study is  to present a detailed and 
systematical study of the above-mentioned problem  
using a fully non-equilibrium formalism~\cite{boy1,boy2,boy3,boy4,lee}. 
We will derive the coupled nonperturbative equation 
for the  axion field and the mode equations for the photon field
in a flat Robertson-Walker spacetime within the nonperturbative Hartree 
approximation that is implemented to  consistently take  the back reaction 
effects into account.  
We then try to study both numerically and analytically  how the nonperturbative 
effects of  spinodal instability and parametric amplification of quantum 
fluctuations trigger photon production from 
the oscillations of the  axion field. At this stage,
it is worthwhile to mention that our approach can be generalized to any 
pseudo-scalar field coupled to the photon field in a more general curved 
spacetime. Since  the pseudo-scalar nature of the coupling between the axion and 
the photon, the  axion field  affects the left- and right-handed circularly 
polarized photons differently. This leads to  producing the two 
polarized photons  in different 
amounts. This polarization asymmetry, if it survives, 
may have  interesting effects on the polarization 
of the cosmic microwave background. 

To consider the fermionic plasma effect on photon production,
one must systematically obtain the non-equilibrium in-medium photon propagators 
and the off-equilibrium effective vertices between the axion and the photon by 
integrating out the fermionic field to deal with this problem\cite{boy4}.
 In a plasma, the transverse photons are 
dynamically screened \cite{bell}.
However, in the literatures~\cite{pre}, the arguments stated to include the 
fermionic plasma effect   in support of
their conclusions amount to adding  by hand the electron plasma frequency  
into the propagating photon mode equations. 
This is problematic when we consider propagating photon modes in the presence of a 
thermal background. In fact, the consequence of  the Abelian ward identities 
reveals   that the transverse photons have the vanishing static magnetic mass in 
all orders of the perturbation theory \cite{bell}. 
This means that  the   in-medium transverse photon propagators 
must be  nonlocal  in nature, and cannot 
be approximated by the local propagator as suggested in Ref.~\cite{pre} 
even in the low energy limit \cite{boy4}. 
Besides, in a
 fermionic plasma, the effective coupling between the axion and the photon 
resulting from integrating out the fermionic thermal loop 
can be modified both at finite temperature as well as out of equilibrium 
\cite{boy4}.
 Therefore,  to fully consider the  plasma effect,  the non-equilibrium in-medium 
photon propagators as well as   the off-equilibrium effective vertices play the 
essential roles. However,  incorporating these non-equilibrium effects   is  a 
challenging  
task that lies beyond the scope of this paper, but certainly deserves 
to be taken up in the near future. 
In the following,  we will  totally ignore the fermionic plasma effect 
in order to focus on understanding
whether the particle production due to spinodal instability as well as parametric 
amplification  is effective or not in the cosmological context.

In Sec.~\ref{AXI}, we introduce the axion physics and its coupling
to the photon. An effective action of the axion-photon system in the 
expanding universe is derived. Sec.~\ref{EQU} is devoted
to the formalism of the problem in terms of  non-equilibrium 
quantum field theory.  We obtain the equation of motion for the classical
axion field and the photon mode equations.
In Sec.~\ref{NUM}, we present the numerical results. Sec.~\ref{CON} is
our conclusions.

\section{EFFECTIVE ACTION  OF  AXION-PHOTON COUPLING IN EXPANDING UNIVERSE}
\label{AXI}

The physics of axion and its implication to astrophysics and cosmology 
can be found in the review articles~\cite{kol,kim}. 
The axion does not couple directly 
to the photon. However, at tree level, the axion field $\phi$ has a 
 coupling with the fermionic field $\psi$,
\begin{equation}
L_{\phi\psi} = g\phi\bar\psi\psi.
\end{equation}
Therefore, the axion can couple to the photon via a fermionic loop. 
As a consequence of the color anomaly of the PQ current, similar to
the pion-photon system, the effective Lagrangian  density for the axion-photon
coupling is
\begin{equation}
L_{\phi A} = c \frac{\phi}{f_a}\epsilon^{\alpha\beta\mu\nu} 
            F_{\alpha\beta} F_{\mu\nu},
\label{L_phiA}
\end{equation}
where $F_{\mu\nu}=\partial_\mu A_\nu - \partial_\nu A_\mu$. 
In Eq.~(\ref{L_phiA}), the scale $f_a\equiv f_{PQ}/N$, where $f_{PQ}$ is the
PQ symmetry breaking scale, and $N$ is the color anomaly of the PQ symmetry.
The coupling constant $c=\alpha(E_{PQ}/N-1.95)/(16\pi)$, where $E_{PQ}$ is
the electromagnetic anomaly of the PQ symmetry, and $\alpha$ is the fine
structure constant. Henceforth, we assume an axion incorporated into the
 simplest GUTs with $E_{PQ}/N=8/3$, such that 
 $ c \simeq 1.04\times 10^{-4}$~\cite{din}.

Now we write down the effective action for the axion-photon system
in the expanding universe,
\begin{equation}
S=\int d^4x {\sqrt g} \left( L_\phi + L_A + \frac{1}{\sqrt g} L_{\phi A}\right),
\label{action}
\end{equation}  
where $1/\sqrt g$  is added in front of $L_{\phi A}$, given by
Eq.~(\ref{L_phiA}), because $\epsilon^{\alpha\beta\mu\nu}$ 
is a tensor density of weight $-1$~\cite{wei}.
In the following, for simplicity, we will assume a flat Robertson-Walker metric
\begin{equation}
ds^2= -g_{\mu\nu} dx^\mu dx^\nu = dt^2- a^2(t) d{\vec x}^2,
\label{metric}
\end{equation}
where the signature is $(-+++)$, and $a(t)$ is the cosmic scale factor.
In Eq.~(\ref{action}), 
\begin{eqnarray}
L_\phi &=&  -{1\over2}g^{\mu\nu}\partial_\mu\phi\partial_\nu\phi - V(\phi,T), \\
L_A &=& -{1\over4} g^{\alpha\mu} g^{\beta\nu} F_{\alpha\beta} F_{\mu\nu},
\end{eqnarray}
where the axion potential has a temperature-dependent mass term due to 
QCD instanton effects, being of the form~\cite{kol}
\begin{eqnarray}
V(\phi,T)&=&m_a^2(T) f_a^2 \left(1-\cos\frac{\phi}{f_a}\right), \\
m_a(T) &\simeq& 0.1 m_{a0} \left(\frac{\Lambda_{QCD}}{T}\right)^{3.7},
\label{maT}
\end{eqnarray}
where $m_{a0}$ is the zero-temperature axion mass, satisfying 
$m_{a0} f_a \simeq 6.2\times 10^{-3} {\rm GeV}^2$. Also, we use
$\Lambda_{QCD}\simeq 200 {\rm MeV}$.

It is well known that the minimal coupling of photons to the metric
background is conformally invariant~\cite{wid}. 
As such, in the conformally flat metric~(\ref{metric}), 
it is convenient to work with the conformal time,
$d\eta=a^{-1}(t)dt$. Hence, defining $\phi=\chi/a$, the action~(\ref{action})
becomes
\begin{eqnarray}
S= \int d\eta ~d^3{\vec x}~~ {\cal{L}}=\int d\eta~ d^3{\vec x} ~ && \left[ 
  {1\over2} \left(\frac{\partial\chi}{\partial\eta}\right)^2 - 
  {1\over2} \left(\frac{\partial\chi}{\partial{\vec x}}\right)^2 +
  {1\over2 a} \frac{d^2 a}{d\eta^2} \chi^2 - a^4 V(\frac{\chi}{a},T)
  \right. \nonumber \\ && \left. 
  -{1\over4} \eta^{\alpha\mu} \eta^{\beta\nu} F_{\alpha\beta} F_{\mu\nu}
  + c \frac{\chi}{af_a}\epsilon^{\alpha\beta\mu\nu} F_{\alpha\beta} F_{\mu\nu}
  \right],
\label{newact}
\end{eqnarray}
where $\eta^{\mu\nu}$ is the Minkowski metric. In terms of the conformal time, the 
effective action now has analogy with the effective action in Minkowski
spacetime with the time dependent mass term and interactions.

\section{ EQUATIONS OF MOTION}
\label{EQU}
The non-equilibrium effective Lagrangian in the  closed time path 
formalism\cite{boy1,boy2,boy3} is given by
\begin{equation}
{\cal L}_{neq}={\cal L}\left[\chi^+,A_\mu^+\right] -
               {\cal L}\left[\chi^-,A_\mu^-\right],
\label{Lneq}
\end{equation}
where + (-) denotes the forward (backward) time branches. 
We then decompose $\chi^{\pm}$ into the axionic mean field and 
the associated quantum fluctuating fields:
\begin{equation}
\chi^{\pm}(\vec x,\eta)=\varphi(\eta) + \psi^{\pm}(\vec x,\eta),
\label{mean}
\end{equation}
with the tadpole conditions,
\begin{equation}
\langle \psi^{\pm}(\vec x,\eta) \rangle = 0.
\label{tad}
\end{equation}
We will implement the  tadpole conditions  to all orders in the corresponding 
expansion to obtain the non-equilibrium equations of motion.

To 
take account of the back reaction effects on the dynamics of the axion field from 
quantum fluctuating photon modes, 
we adopt the following  
Hartree factorization  
which is  implemented for both $\pm$ components\cite{boy1,boy2,boy3}:
\begin{equation}
 \psi F {\tilde{F}}  \rightarrow \psi \langle F {\tilde{F}} \rangle.
\label{har} 
\end{equation}
As seen later, the expectation value can be determined self-consistently. It must 
be noted   that there is no {\it a priori} justification for such a factorization. 
However, this approximation provides a nonperturbative framework that allows us to 
treat photon fluctuations self-consistently~\cite{boy3} . 
On the contrary,   we will ignore the quantum fluctuations of the axion, which can 
be
produced via self-couplings, as the study of Ref.~\cite{kol2} has shown that these 
effects are insignificant.

With Eq.~(\ref{mean}), we first expand the non-equilibrium Lagrangian density  
(\ref{Lneq}) in powers of $\psi$ and keep the 
term up to  linear   $\psi$  to  ignore its quantum fluctuation effects.    
 Together with Eq.~(\ref{har}), the Hartree-factorized Lagrangian then becomes
\begin{eqnarray}
{\cal L}\left[\varphi(\eta)+\psi^+,A_\mu^+\right] -
               {\cal L}\left[\varphi(\eta)+\psi^-,A_\mu^-\right] 
&& =\left\{ - U (\eta) \psi^+ 
            -{1\over4}F^{+}_{\mu\nu}{\tilde{F}}^{+\mu\nu} 
     +\frac{c}{a f_a} \varphi(\eta) F^+_{\mu\nu} {\tilde{F}}^{+\mu\nu} \right. 
\nonumber \\
&& \left.
    +\frac{c}{a f_a} \psi^+  \langle F^{+\mu\nu}{\tilde{F}}^+_{\mu\nu} \rangle
         \right\}  
-\left\{ + \rightarrow -\right\}, 
\end{eqnarray} 
where 
\begin{eqnarray}
 U(\eta) &=&  \ddot\varphi(\eta) -\frac{ \ddot{a} (\eta)}{ a(\eta)} \varphi(\eta) 
+  a^3 (\eta)~ m_a^{2}(T) f_a \sin 
\left[\frac{\varphi (\eta)}{ a(\eta) f_a}\right]\,.  
\end{eqnarray}
The dot means the time derivative with respect to the conformal time.

With the tadpole conditions~(\ref{tad}), 
we obtain  the following  equation of motion 
for the axionic mean field given by  
  \begin{equation}
\ddot\theta(\eta)+ 2\frac{{\dot{a}}(\eta)}{a(\eta)} \dot\theta(\eta)
          +  a^2 (\eta)~ m_a^{2} (T)  \sin\theta (\eta) -  \frac{1}{ a^2(\eta) }
     \left( \frac{c}{ f^2_a} \right) \langle F {\tilde{F}} \rangle (\eta) =0 \,, 
\label{eom}
\end{equation}
where we define the dimensionless field, $ \theta (\eta) = \varphi (\eta) 
/(a(\eta) f_a) $.

Within the Hartree approximation, the photon production processes  do not involve 
photons in the intermediate states~\cite{boy3,lee}. To avoid the gauge 
ambiguities, we will work in the coulomb gauge and concentrate only on physical 
transverse gauge field, $ {\vec{A}}_T (\vec x,\eta)$  \cite{boy3,lee}. 
Then, the Heisenberg field equation for $ {\vec{A}}_T (\vec x,\eta)$    
can be read off from the quadratic part 
of the Lagrangian in the form
\begin{equation}
\frac{d^2}{d\eta^2} {\vec A_T}(\vec x,\eta) -
\vec\nabla^2\vec A_T(\vec x,\eta)+4 c~ \dot\theta (\eta) \vec\nabla \times 
\vec A_T (\vec x,\eta) =0. 
\end{equation}
It is more convenient to  decompose the field 
$\vec A_T (\vec x,\eta)$ into the Fourier
mode functions  $V_{\lambda \vec k}(\eta)$ in terms of circularly polarized 
states,
\begin{eqnarray}
\vec A_T(\vec x,\eta) &=&\int
\frac{d^3k}{\sqrt{2(2\pi)^3 k}} \vec A_T(\vec k,\eta) \nonumber \\
&=& \int
\frac{d^3k}{\sqrt{2(2\pi)^3 k}}
\left\{ \left[b_{+ \vec k} V_{1 \vec k}(\eta) \vec \epsilon_{+ \vec k}+b_{- \vec 
k} 
V_{2 \vec k}(\eta) \vec \epsilon_{- \vec k}\right]
 e^{i\vec k\cdot \vec x} +
{\rm h.c.}\right\},
\end{eqnarray}
where $b_{\pm \vec k}$ are destruction operators, and
$\vec \epsilon_{\pm \vec k}$ are circular polarization unit vectors defined in 
Ref.~\cite{boy3}.
Then the mode equations are
\begin{eqnarray}
&&\frac{d^2 V_{1 k}(\eta)}{d\eta^2} + k^2 V_{1 k}(\eta)-4k \,c\, \dot{\theta} 
(\eta) V_{1 k}(\eta)=0,  \nonumber \\
&& \frac{d^2 V_{2 k}(\eta)}{d\eta^2} + k^2 V_{2 k}(\eta)+4k\, c\, \dot{\theta} 
(\eta) V_{2 k}(\eta)=0,
\label{meq}
\end{eqnarray}
with the  expectation values given by \cite{boy3}
\begin{equation}
\langle F \tilde{F} \rangle(\eta)=
\frac{1}{\pi^2} \int k^2 dk \coth \left[ \frac{k}{2 T_i} \right]
\frac{d}{d\eta}
\left( |V_{1 k}(\eta)|^2- |V_{2 k}(\eta)|^2 \right),
\label{qf}
\end{equation} 
where we have assumed that at initial time $\eta_i$, the photons are in local 
equilibrium at the initial temperature $T_i$. Clearly, the photon mode equations 
(\ref{meq})
are decoupled in terms of the circular polarization mode functions. The axion 
field acts as the time dependent  mass term thats triggers photon production.  
The effective mass terms have opposite signs for the 
two  polarizations due to  the pseudo-scalar nature of the axion-photon
coupling. This will lead to producing the different polarized photons in different 
amounts, resulting in a polarization asymmetry in photon emission.  
The expectation
value of the number operator for the asymptotic photons with momentum 
$\vec k$ is given by\cite{boy3}
\begin{eqnarray}
\langle {\bf N}_{k}(\eta)\rangle&=&
 {\frac{1}{2k}} \left[ \dot{\vec A_T}( {\vec k},\eta) \cdot 
\dot{\vec A_T}( -{\vec k},\eta) \right. \left.
+k^2  {\vec A_T}( {\vec k},\eta) \cdot 
{\vec A_T}( {-\vec k},\eta) \right]-1 \nonumber \\  
&=& {1\over 4k^2} \coth \left[ \frac{k}{2 T_i} \right]
\left[|\dot V_{1 k}(\eta)|^2+k^2
| V_{1 k}(\eta)|^2\right] - \frac{1}{2}+ \nonumber \\
&& {1\over 4k^2} \coth \left[ \frac{k}{2 T_i} \right]
\left[|\dot V_{2 k}(\eta)|^2+k^2
| V_{2 k}(\eta)|^2\right] - \frac{1}{2} \nonumber \\
&=  &  N_{+}(k,\eta) +N_{-}(k,\eta),
\label{numop}
\end{eqnarray}
which is the number of photons with momentum 
$\vec k$ per unit comoving volume.

\section{NUMERICAL RESULTS}
\label{NUM}

In this section, we will compute the photon production from the axionic 
condensate with $f_a=10^{12}{\rm GeV}$, which has a zero-temperature
mass $m_{a0}\simeq 6.2\times 10^{-6}{\rm eV}$. As we will show below,
the photon production takes place mainly during the radiation-dominated 
epoch. So, we simply assume a radiation-dominated universe.

At temperature $T$, the Hubble parameter is
\begin{equation}
H\equiv \left(\frac{\dot a}{a}\right)^2 = 
{5\over3} g^{1\over 2}(T) \frac{T^2}{m_{pl}},
\label{Hubble}
\end{equation}
where $g(T)$ is the number of effective degrees of freedom at temperature $T$, 
$m_{pl}$ is the Planck scale, and the cosmic scale factor is  
\begin{equation}
a(\eta)=\frac{\eta}{\eta_1},
\end{equation}
where $\eta_1$ is the time when the axion field starts to oscillate, defined by
a temperature $T_1$ such that $3H(T_1)=m_a(T_1)$. As such, 
$\eta_1^{-1}=m_a(T_1)/3$. For $f_a=10^{12}{\rm GeV}$,
from Eqs.~(\ref{maT}) and (\ref{Hubble}) we find $T_1\simeq 0.9\,{\rm GeV}$
and $g(T_1)\simeq 60$.

Changing the variable $\eta$ into $a$, the equations of motion~(\ref{eom})
and (\ref{meq}) become
\begin{eqnarray}
&&\frac{d^2\theta}{da^2}+{2\over a}\frac{d\theta}{da}
      +  9a^2 \frac{m_a^2(T)}{m_a^2(T_1)}\sin\theta
      -\frac{9c}{m_a^2(T_1) a^2 f^2_a}~ \langle F {\tilde{F}} \rangle =0 \,, 
\label{eom2}  \\
&& \frac{d^2 V_{1 \xi}}{da^2} + \xi^2 V_{1 \xi} 
      -4\xi \,c\, \frac{d\theta}{da} V_{1 \xi} =0,  \nonumber \\
&& \frac{d^2 V_{2 \xi}}{da^2} + \xi^2 V_{2 \xi}
      +4\xi \,c\, \frac{d\theta}{da} V_{2 \xi} =0,
\label{meq2}
\end{eqnarray}
where $\xi=k\eta_1$ and 
\begin{equation}
\frac{m_a^2(T)}{m_a^2(T_1)}=\left\{
\begin{array}{ll}
\left({T_1\over T}\right)^{7.4}=a^{7.4}\left[\frac{g(T)}{g(T_1)} 
\right]^{7.4\over 4} & T >> \Lambda_{QCD}, \\
10^2 \left(\frac{T_1}{\Lambda_{QCD}}\right)^{7.4} & T << \Lambda_{QCD}.
\end{array}
\right.
\label{mass}
\end{equation}
Note that $g(T)$ does not change significantly from $T_1$ to $\Lambda_{QCD}$.
Henceforth, we approximate $g(T)\simeq g(T_1)\simeq 60$, where 
$T_1= 0.9\,{\rm GeV}$. It is worth to point out that the mode 
equations~(\ref{meq2}) have unstable modes via the spinodal
instability for sufficiently low-momentum modes with $\xi< 4c|d\theta/da|$,
where the effective mass becomes negative.  

To solve Eqs.~(\ref{eom2}) and (\ref{meq2}), we have to specify
the initial conditions for the axion and photon fields. The amplitude of 
the axion field is frozen for $\eta << \eta_1$, i.e.,
\begin{equation}
\theta=1,\quad \frac{d\theta}{da}=0,\quad {\rm as}\;\;a=a_i << 1.
\end{equation}
For the photon mode functions, we propose
\begin{equation}
V_{1 \xi}=V_{2 \xi}=1,\quad \frac{dV_{1 \xi}}{da}=\frac{dV_{2 \xi}}{da}
=-i\xi,\quad {\rm as}\;\;a=a_i << 1.
\end{equation}
These initial conditions are physically plausible and simple enough for us
to investigate a quantitative description of the dynamics. 
To evaluate the $\langle F \tilde{F} \rangle$ in Eq.~(\ref{qf})
and the photon number operator in Eq.~(\ref{numop}), we approximate
the bose enhancement factor by
\begin{equation}
\coth \left[ \frac{k}{2 T_i} \right] = \coth \left[ \frac{\xi}{2\Gamma} \right]
\simeq \frac{2\Gamma}{\xi},
\end{equation}
where $\Gamma\equiv \eta_1 T_i \ge \eta_1 T_1 \simeq 10^{18}$, 
and we are interested in $\xi < 100$.

In Fig.~\ref{fig1}, we plot the temporal evolution of the axion field and 
its time 
derivative by choosing $a_i =0.01$ that corresponds to $T_i=100\,T_1$. 
Due to the expansion of the universe, the field amplitude decreases
with time. However, the rate of change of the amplitude increases with time.
To understand this, we redefine $\theta\equiv \tilde{\theta}/a$ 
in Eq.~(\ref{eom2}).  The $\langle F \tilde{F} \rangle$ term can be neglected,
being extremely small as shown in Fig.~\ref{fig2}, where we have evaluated
the last term of Eq.~(\ref{eom2}) denoted by $\Sigma (a)$. 
Then, the equation of motion for $\tilde{\theta}$ when $\theta << 1$ is given by 
\begin{equation}
\frac{d^2 \tilde{\theta}}{da^2} + 9a^2 \frac{m_a^2(T)}{m_a^2(T_1)}
\tilde{\theta} =0.
\label{eom3}
\end{equation}
From Eq.~(\ref{mass}), the solutions for $\tilde{\theta}$ are Bessel functions.
For $T>>\Lambda_{QCD}$, asymptotically 
$\tilde{\theta} \propto \cos(0.53\,a^{5.7})/a^{2.35}$.
For $T<<\Lambda_{QCD}$, $\tilde{\theta} \propto \cos(3917\,a^2)/a^{0.5}$.
The latter shows that the axions behave like non-relativistic matter.

For the photon production, we calculate the spectral photon number density
$N(\xi,a)$, equal to $\langle {\bf N}_{k}(\eta)\rangle$
in Eq.~(\ref{numop}). It is convenient to define a ratio,
\begin{equation}
n(\xi,a)\equiv \frac{N(\xi,a)-N(\xi,a_i)}{N(\xi,a_i)},
\label{ratio}
\end{equation}
which is the excess photons above the
thermal background. Three snapshots of the ratio at $a=0.75, 1.5,$ and $3$ are 
shown in Fig.~\ref{fig3}. 
It is interesting to see that the production duration
of each non-zero mode is short, and higher-momentum modes are produced
at later times. This in fact demonstrates a brief exponential growth of
the unstable mode due to the parametric resonance instability, which is 
shut-off when the unstable mode has been red-shifted  
out of the unstable band. As a consequence, the photon production is 
limited with an excess photon ratio typically at a level of $10^{-7}$.
As we have mentioned above, the spinodal instability happens for low-momentum
photon modes. We demonstrate this numerically in Fig.~\ref{fig4}, 
where we have chosen
$\xi<0.004$ that lie within the spinoidal region where $\xi< 4c|d\theta/da|$. The 
production ratio is also at a level of $10^{-7}$, but it is
apparent that these low-momentum modes are produced during the first
oscillating cycle of the axion field.

The ratio in Fig.~\ref{fig3} can be actually estimated as follows. 
First, let us find out 
the approximate form for $d\theta/da$ from Eq.~(\ref{eom3}) 
and Fig.~\ref{fig1}, which is given by
\begin{equation}
\frac{d\theta}{da}\simeq 3.6\,a^{1.35} \cos(0.53\,a^{5.7})
\quad {\rm for} \quad  a_i << a< 8.4,
\label{dtheta}
\end{equation}
where $a=8.4$ is the scale factor when $T=\Lambda_{QCD}$. At instant $a$, 
$d\theta/da$ is oscillating with an effective frequency $\omega=0.53\,a^{4.7}$.
Inserting the approximate form~(\ref{dtheta}) 
with $\omega$ treated as a constant into 
the mode equation of $V_{1 \xi}$ in Eq.~(\ref{meq2}), and changing 
variable to $z=\omega a/2$, we have 
\begin{equation}
 \frac{d^2 V_{1 \xi}}{dz^2} + \frac{4\xi^2}{\omega^2} V_{1 \xi} 
      - 58.1\,c \frac{\xi}{\omega^2} a^{1.35} \cos(2z) V_{1 \xi} =0.
\end{equation}
This is the standard Mathieu equations~\cite{mac}. 
The widest and most important instability is the first parametric resonance 
that occurs at $\xi=\omega/2$ with a narrow
bandwidth $\delta\simeq 14.5\, c \,a^{1.35}/\omega$.
But actually $\omega$ is changing with time. As such, the unstable mode 
will grow exponentially only during a brief period roughly given by
\begin{equation}
\Delta z \simeq \frac{\omega \delta}{\frac{\Delta\omega}{\Delta z}}
\simeq \frac{\omega^2 \delta}{2\frac{\Delta\omega}{\Delta a}}.
\end{equation}
Consequently, this instability leads to the growth of the occupation numbers 
of the created photons by a growth factor,
\begin{equation}
e^{2\mu\Delta z}\simeq 1+2\mu\Delta z, \nonumber
\end{equation}
where the growth index $\mu\simeq \delta/2$. Hence the mode number density
is increased by an amount approximately given by
\begin{equation}
2\mu\Delta z \simeq 42.6\, c^2 a^{-1} \simeq 4.6\times 10^{-7} a^{-1}, \nonumber
\end{equation}
where $a$ is the scale factor when the mode $V_{1 \xi}$ enters into 
the resonance band. This estimation is of the same order of magnitude 
as found in the numerical results. Interestingly, the $a^{-1}$ 
dependence of the mode production can also be seen in Fig.~\ref{fig3}. 
Similar estimation can also be done for the mode function $V_{2 \xi}$.

The polarization asymmetry in the produced photons is defined as
\begin{equation}
\Xi (\xi,a) =\frac{ N_{+}(\xi,a) -N_{-}(\xi,a) }
{ N_{+}(\xi,a) + N_{-}(\xi,a) }.
\label{asymm}
\end{equation}
We have input the mode solutions for 
$V_{1 \xi}$ and $V_{2 \xi}$ to Eq.~(\ref{numop}) to
calculate  $\Xi(\xi,a)$~(\ref{asymm}). 
A plot of the asymmetry versus the momentum at $a=3$ 
is shown in Fig.~\ref{fig5}. 
The numerical result shows that the asymmetry is fluctuating about zero
as the photon momentum varies. The fluctuating amplitude is about
$10^{-8}$ of the thermal background. Although the asymmetry averaged out over
a wide range of momenta is nearly zero, at certain momenta
the produced photons are about $10\%$ circularly polarized. 
However, subsequently this polarization asymmetry 
will be damped out by photon-electron scatterings in the plasma.

\section{CONCLUSIONS}
\label{CON}

First we basically confirm   that the photon 
production via 
parametric resonance is invalidated by the expansion of the universe. 
Besides, we find that in addition to parametric
resonance, for long-wavelength photon modes, a new dissipative channel
via spinodal instability is open. This open channel results in the long-wavelength 
fluctuations of the photon modes. Again, this production is suppressed in the
expanding universe. We also observe the polarization asymmetry in 
the produced circularly polarized photons as a result of the pseudo-scalar nature 
of the coupling. However, it will be damped out effectively by the plasma. 
But it is very interesting to see whether it is possible to generate a circular
polarization asymmetry in the production of photons from 
certain pseudo-scalar fields such that
it may leave an imprint on the polarization of the cosmic microwave background. 
As to the plasma damping on photon production,
we have pointed out
the problem in the naive approximation that simply introduces the electron plasma 
frequency 
to the photon modes. We have thus proposed a dynamical and non-equilibrium 
 treatment which should be a better approach to consider the plasma effects.
The actual calculations are rather difficult, 
but they certainly deserve further studies.

\acknowledgments

We would like to thank D. Boyanovsky and H. J. de Vega for their useful 
discussions. 
The work of D.S.L. (K.W.N.) was supported in part by the 
National Science Council, ROC under the Grant 
NSC89-2112-M-259-008-Y (NSC89-2112-M-001-001).

\newpage

\begin{center}
{\bf FIGURE CAPTIONS}
\end{center}
\bigskip

\begin{figure}
\leavevmode
\hbox{
\epsfxsize=5.5in
\epsffile{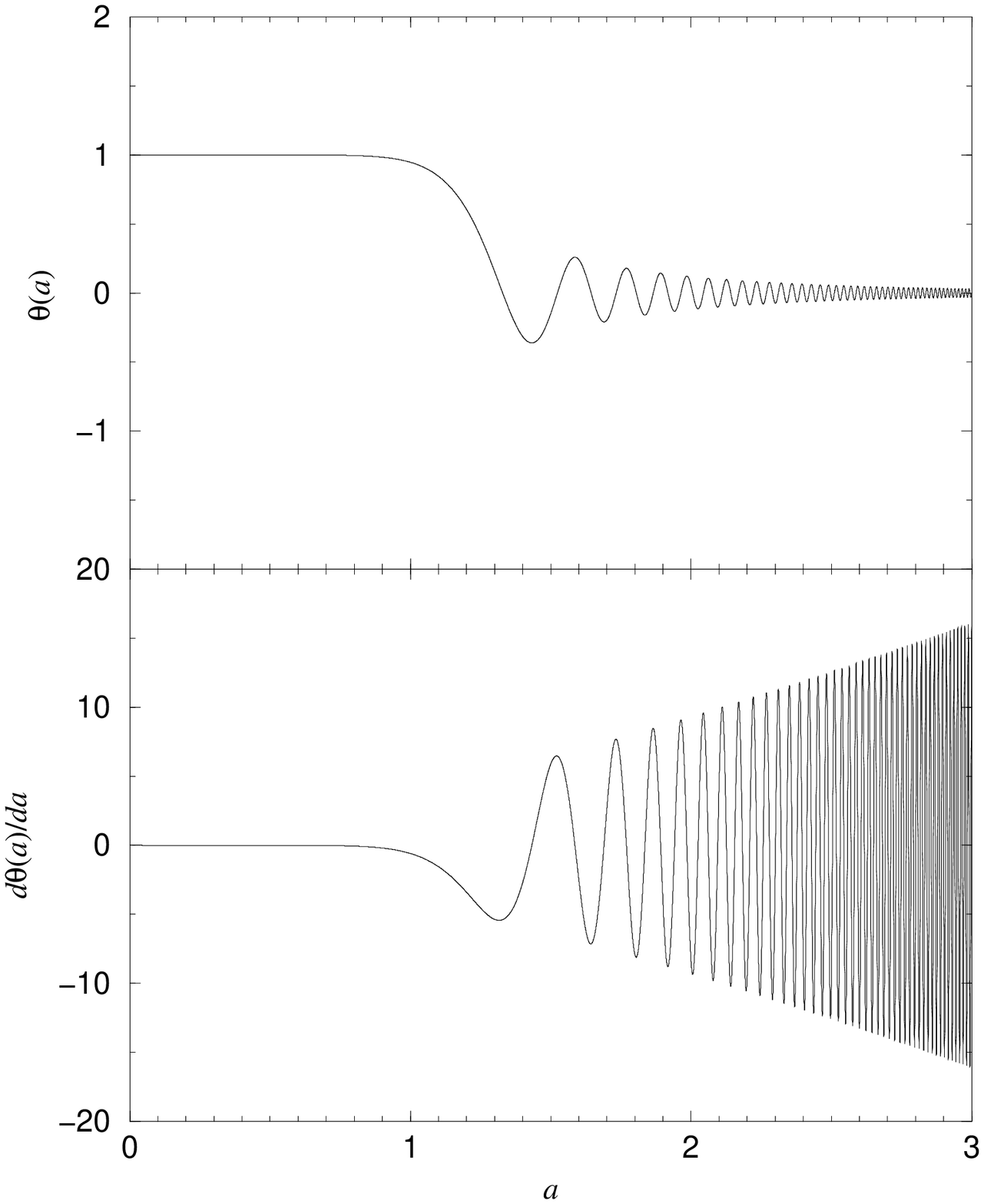}}
\caption{}
\label{fig1}
\end{figure}

\noindent
Fig.~\ref{fig1}. 
Time evolution of the axion mean field $\theta(a)$ and its time derivative
$d\theta(a)/da$, where $a$ is the cosmic scale factor.
\medskip

\begin{figure}
\leavevmode
\hbox{
\epsfxsize=5.5in
\epsffile{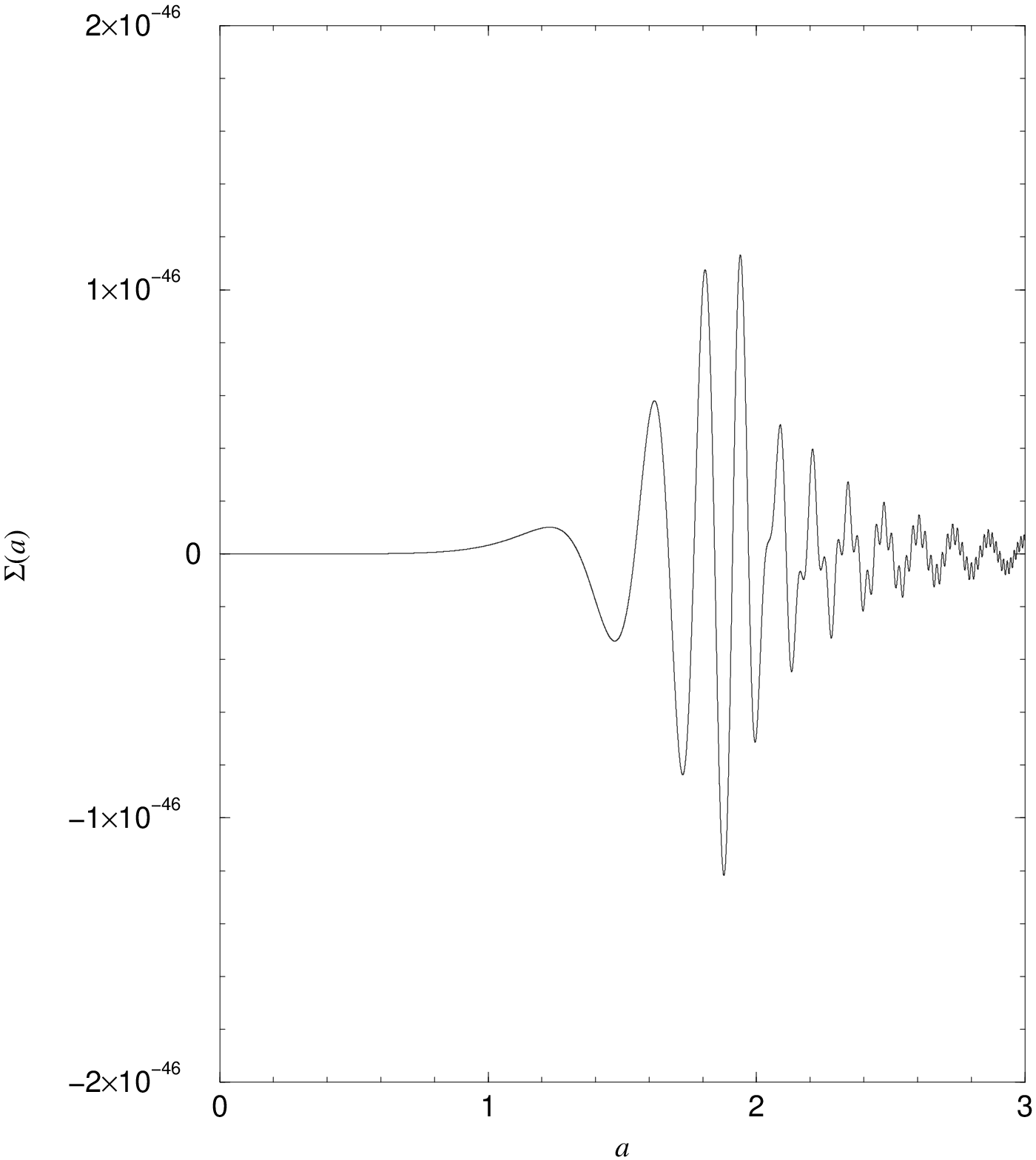}}
\caption{}
\label{fig2}
\end{figure}

\noindent
Fig.~\ref{fig2}. 
Time evolution of $\langle F \tilde{F} \rangle$, plotted with the
quantity $\Sigma (a)$ given by the last term of Eq.~(\ref{eom2}).
\medskip

\begin{figure}
\leavevmode
\hbox{
\epsfxsize=5.5in
\epsffile{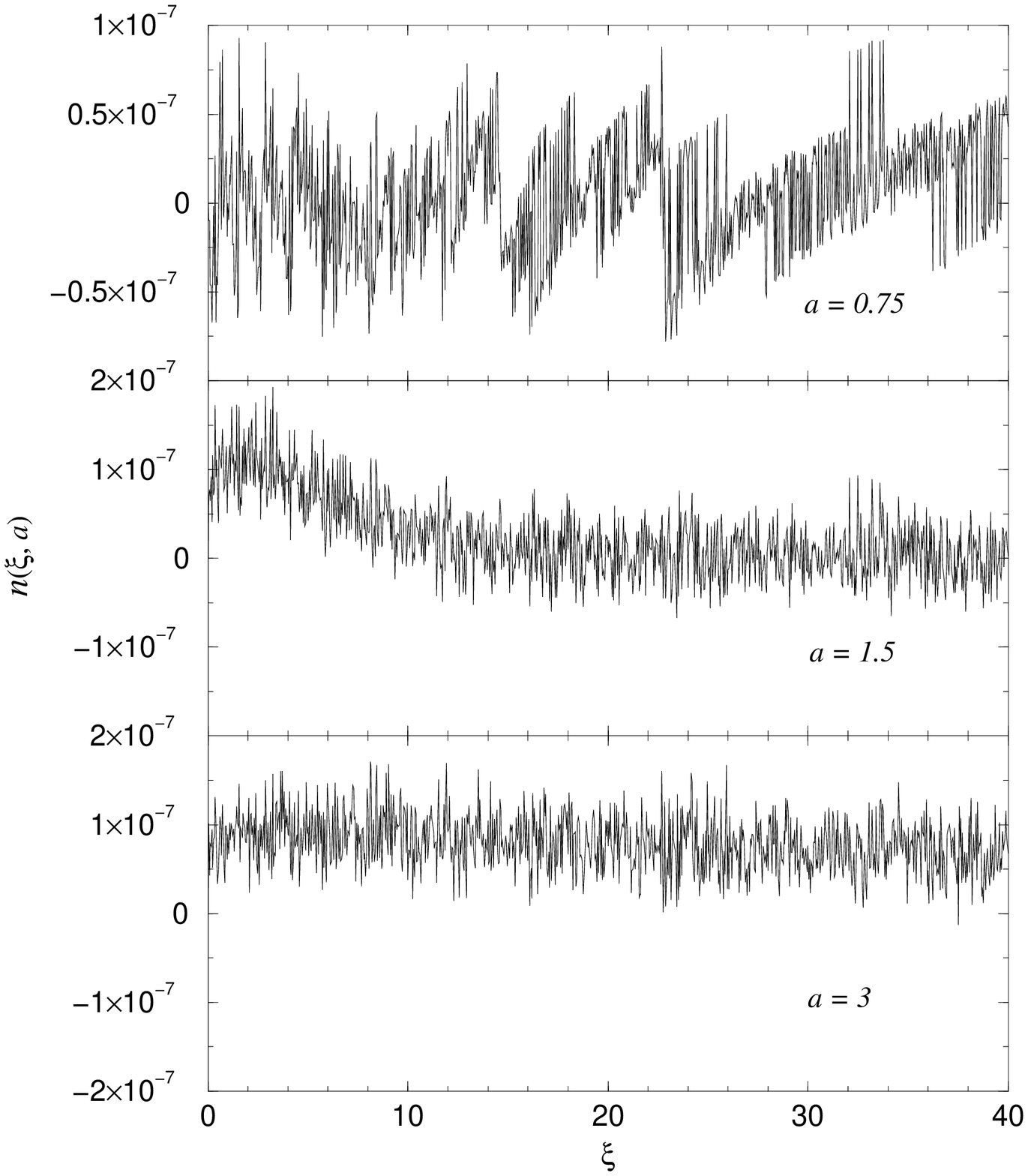}}
\caption{}
\label{fig3}
\end{figure}

\noindent
Fig.~\ref{fig3}. 
Three snapshots of the spectral number density ratio $n(\xi,a)$, defined
in Eq.~(\ref{ratio}), of the photons produced via parametric
amplification at $a=0.75, 1.5,$ and $3$. The dimensionless quantity
$\xi=k\eta_1$, where $k$ is the photon momentum and 
$\eta_1$ is the conformal time when the axion field starts to oscillate.
\medskip

\begin{figure}
\leavevmode
\hbox{
\epsfxsize=5.5in
\epsffile{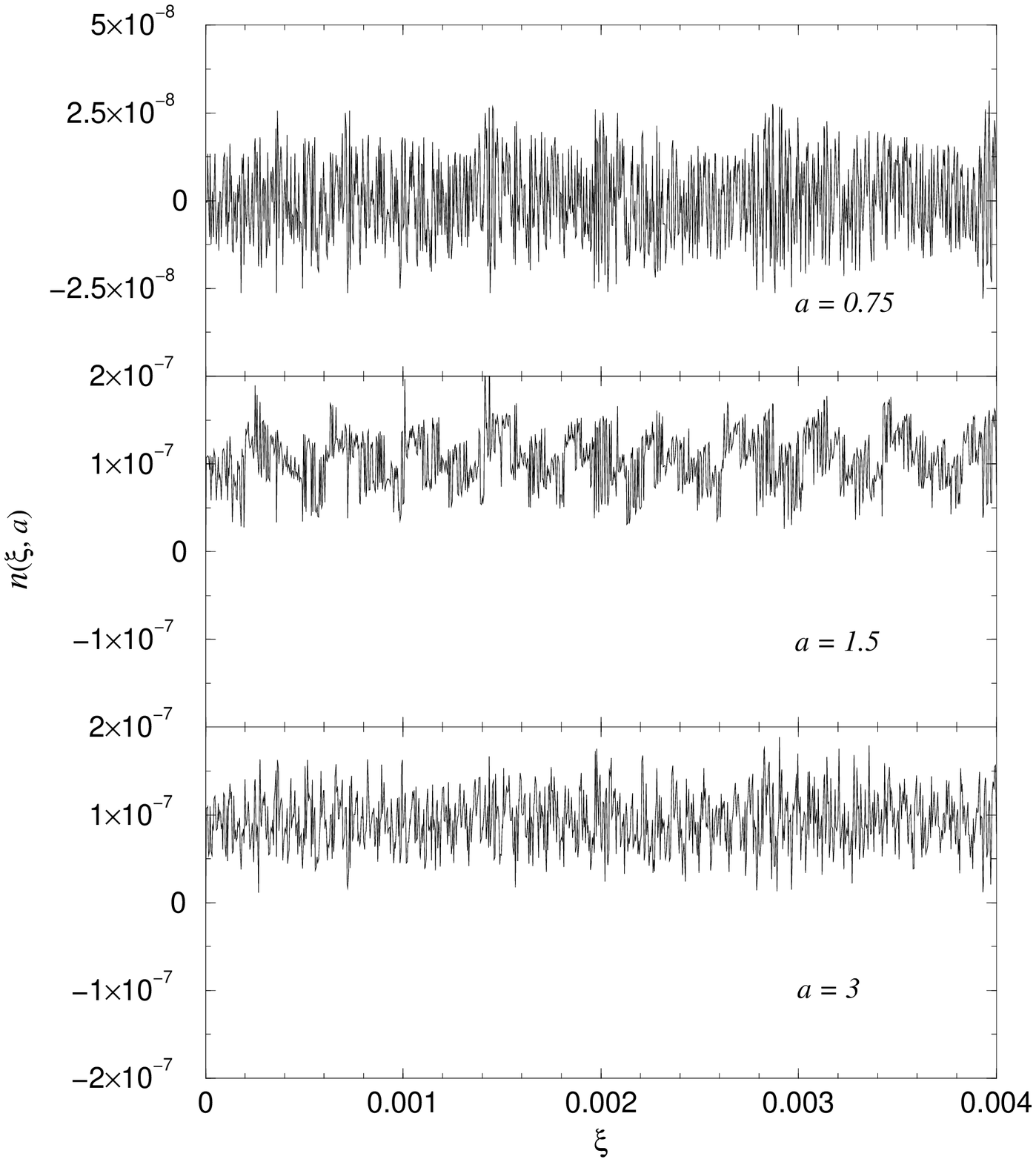}}
\caption{}
\label{fig4}
\end{figure}

\noindent
Fig.~\ref{fig4}. 
As in Fig.~\ref{fig3}, but for low-momentum photons produced via 
spinodal instability. 
\medskip

\begin{figure}
\leavevmode
\hbox{
\epsfxsize=5.5in
\epsffile{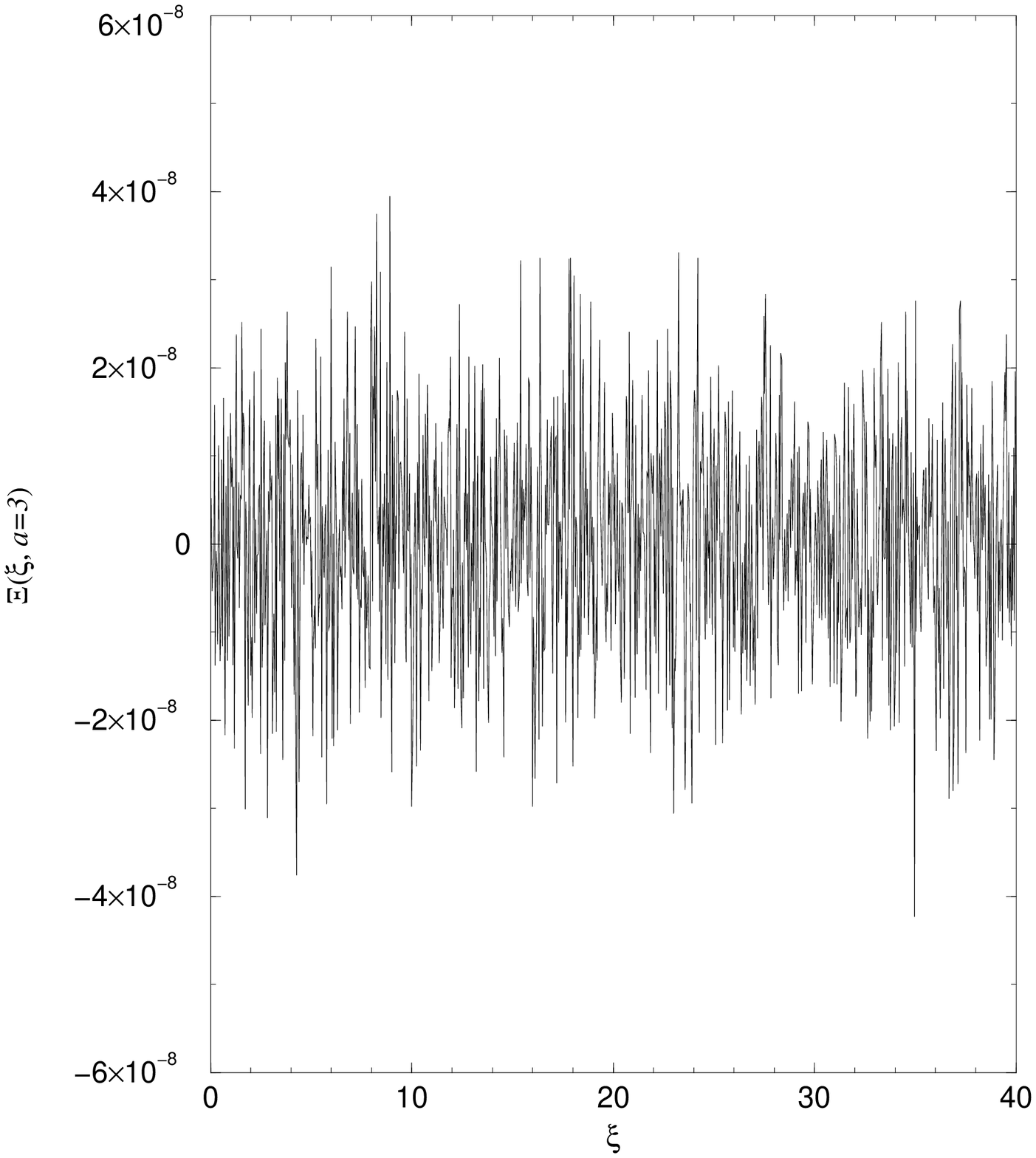}}
\caption{}
\label{fig5}
\end{figure}

\noindent
Fig.~\ref{fig5}. 
Circular polarization asymmetry $\Xi (\xi,a)$, defined in Eq.~(\ref{asymm}),
of the photons at $a=3$ for $0<\xi<40$.

\newpage

\newpage

\newpage

\newpage

\newpage

\end{document}